\documentclass{article}

\usepackage{PRIMEarxiv}

\usepackage[utf8]{inputenc} 
\usepackage[T1]{fontenc}    
\usepackage{hyperref}       
\usepackage{url}            
\usepackage{booktabs}       
\usepackage{amsfonts}       
\usepackage{nicefrac}       
\usepackage{microtype}      
\usepackage{lipsum}
\usepackage{graphicx}
\graphicspath{{media/}}     

\title{A new data weighted averaging algorithm to reduce tones in the signal band}

\author{
  Marta Laguna, Juana M. Martínez-Heredia \\
  Departamento de Ingeniería Electrónica \\
  Universidad de Sevilla \\
  Sevilla, Spain\\
  \texttt{\{martalaguna, jmmh\}@us.es} \\
   \And
  Manuel G. Satué \\
  Departamento de Ingeniería de Sistemas y Automática \\
  Universidad de Sevilla \\
  Sevilla, Spain\\
  \texttt{mgarrido16@us.es} \\
}

\begin{document}
\maketitle

\begin{abstract}
Digital/Analog converters based on sigma-delta modulation  are simple and unexpensive circuits featuring a signal bandwidth  limited by speed constraints. Multi-bit modulators allow balancing complexity and speed by reducing the clock frequency and increasing the number of levels in the  quantizer. In this case, the multi-bit digital to analog block (DAC) can reduce the performance of the entire system. Data Weighted Averaging (DWA) methods have been proposed to reduce the vulnerability to DAC errors at the cost of spurious tones in the signal band. This work analyzes the tone producing mechanism and proposes a modification of the DWA to remove spurious tones.
\end{abstract}


\section{Introduction}
Sigma Delta Modulators (SDM) are used in analog/digital converters \cite{candy1991oversampling}, specially in equipment using Complementary Metal-Oxide-Semiconductor devices \cite{colodro2005time}. One particular technique in SDM design is multibit quantizer \cite{leung2002multibit,lindfors2002two}. They are placed in the modulator control path providing improved bandwidth. Multibit SDM have the advantage of retaining resolution for low oversampling situations \cite{yang2024research}. Nonlinearities in voltage-controlled oscillator quantizers is proposed in this brief. The signal to be converted have been analyzed in \cite{colodro2014linearity} resulting in methods to reduce harmonic distortion. Another advantage of multibit SDM is that each quantizer bit provides  6 dB improvement in bandwidth.

However, the internal multibit analog/digital converter needed still requires a high accuracy in the ditital to analog  conversion (DAC) \cite{colodro2010continuous,kosonocky2017analog}. To overcome this problem, Dynamic Element Matching (DEM)  has been proposed \cite{leung2002multibit}. One particular  DEM, called Data Weighted Averaging (DWA), is specially interesting  for its simplicity \cite{baird1995improved,verreault2024oversampling,
colodro2022time,yuan2023random}

However, the DWA method generate spurious tones at certain input amplitudes \cite{chen1998some,xing202525}. For tones within the signal band can pass through the filter and significantly degrade the performance of the DAC \cite{colodro2003multirate}. To reduce this effect, the DWA method can be changed to select random samples of the SMD signal \cite{baird1995improved}. This can affect the stability, signal to noise ratio and dynamic range. Using more than one pointer in the randomized process has been used in variour proposals such as:  BiDWA \cite{fujimori200090}, Mod-DWA \cite{celin2016reduced} and Segmented-DWA \cite{zhang2007segmented}. All these variants increase the complexity of the simple DWA method.

A simple solution is proposed in \cite{vadipour2002techniques}, where a DC offset is added at the SDM input, making the tones to dissappear. The method however is sensible to the DC offset present in the signal \cite{colodro2024analysis}. To overcome this, this study portraits the use of a binary signal $s(n)$ to be added to the SMD samples $y(n)$ so that the pointer is randomized while removing  spurious tones for any value of DC offset in the input signal, $x(n)$. To do so, an additional element is needed in the DAC to accommodate the sequence $s(n)$ as in \cite{chen1999improved,kumar2022analysis}. 

This work proposes a new technique for selecting DAC elements. A binary signal $s(n)$  is added to $y(n)$, so that the input to the DWA block is $y(n)+s(n)$. The function of $s(n)$ is to randomize the pointer and remove spurious tones for any value of DC offset in the input signal, $x(n)$. As in previous proposals, it is necessary to have an additional element in the DAC to accommodate the sequence $s(n)$ \cite{colodro2009analog}. Next, the tone generation mechanism is reviewed.

\section{Mechanism of tone generation}

To analyze the tone generation process, some simulation results are presented. In these the DWA method works by selecting DAC elements in a circular, successive manner. The signal $y(n)$, is binary encoded from $0$ to $L$, the DAC must have $L$ elements, with gains $g_k$, for $k=1, ..., L$, of which $y(n)$ must be selected at the $n$-th clock cycle. The selected elements range from $p(n)$ to $R_L$ $( p(n)+y(n) -1)$ where $p(n)$ is a pointer. $R(k)$ is defined as an integer in the range $1, ..., L$, resulting in

\begin{equation}
\label{eq2_RL}
R_L (k)=k  - L \lfloor (k-1)/L \rfloor
\end{equation}

From this, the next value of the pointer is selected as  $p(n+1)= R_L ( p(n)+ y(n)  )$. Spurious tones are generated by Data Weighted Averaging because the pointer value $p$ is signal dependent, the sequence can feature a limit cycle for certain  $y(n)$ realizations. Then, if the DAC elements do not match, a periodic error may occur. 

In the simulations, a second order SDM with $m=3$ bits is used, resulting in a DAC with $L=7$. To remove tones a new element is introduced yielding a total of $(L+1)=8$ elements. The DAC gains are obtained from a uniformly distributed random generator, resulting in 1.0109, 1.0141, 0.9871, 1.0143, 1.0046, 0.9861, 0.9923, and 1.0016. This produces a standard deviation of 1.16 \%. The sampling frequency is $f_S = 12.5$ (MHz), and the input signal $x(n)$ is a sine function with $f_X = 5.72$ (kHz) and an amplitude of -50 dB full scale. 

\begin{figure}
  \centering  
  \includegraphics[width=14cm]{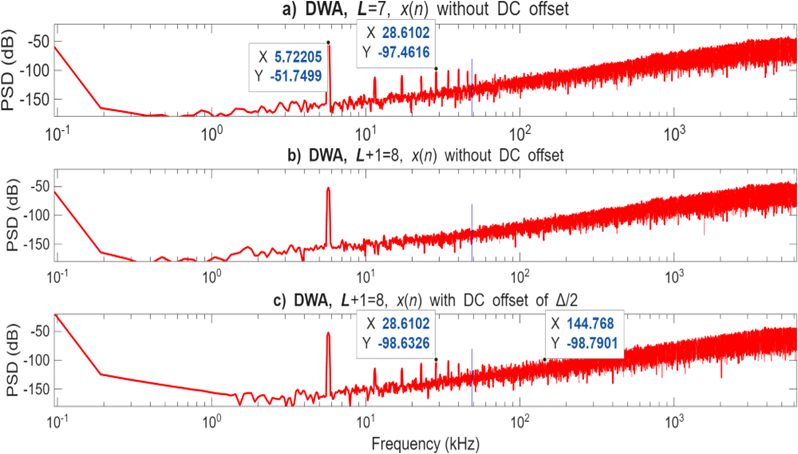}
  \caption{PSD of $v(n)$ for $L$  and $L+1$ DAC elements and for $x(n)$ DC offset $Xoff =0$ and $s(n) =0$.}
  \label{fig_2}
\end{figure}

The $v(n)$ Power Spectral Density (PSD) for the $L$ and $(L+1)$ cases are shown in Fig. \ref{fig_2}. The  SNDR should be 61.7 dB, but just 41.47 dB is measured at the top graph for a signal bandwidth $B = 48.8$ (kHz) (corresponding to an OSR of 128). The reduction in SNDR is due to the large number of tones in the signal band. Observe that, for an amplitude as low as -50 dB and with no DC input offset, the SDM output $y_{sd}(n)$ oscillates around zero taking the values $-\Delta/2$ and $\Delta/2$.

This results in a sequence entering the  DWA of the form $y(n) = 3, 4, 3, 4,...$. The sequence of DAC error time is thus  $\tau(n)=1, 4, 1, 4,...$, producing an error  sequence $e(n)= e_1, e_4, e_1, e_4,...$. This means tha the DAC errors $e(n)$ are not  uniformly distributed values from the set $\{ e_1, ..., e_L \}$ generating harmonic distortion.

The PSD for the  $(L+1)$ case is presented  in the middle on Fig. \ref{fig_2}. The SNDR is 59.07 dB, with no spurious tones and almost full SDNR recovery. This result can be lost if  the signal has its own DC offset. To illustrate this, consider a signal $x(n)$ sinusoidal plus a DC offset of $\Delta/2$. The resulting $y_{sd}(n)$ is approximately equal to $\Delta/2$. In this case $y(n) = 4, 4, ...$ and $\tau(n) = 2, 6, 2, 6,...$ are observed, resulting in $e(n)= e2, e6, e2,...$. The PSD is shown in the lower graph of Fig. \ref{fig_2} where a SDNR of 43.87 dB is found.

\section{Proposed Method}
The proposal revolves around the added sequence concept, where a binary sequence $s(n)$ is added to the DWA input while the  DAC uses one extra element.

In the analysis, the  SDM quantizer has 3-bit resolution. Figure 4 shows the dynamic range curves for a few interesting cases with no DC offset and DAC errors are the given above (standard deviation of 1.16 \%). The obtained dynamic range  is 104 dB,  difference between the input amplitude when SNDR reaches the maximum, -1dB, and the corresponding to SNDR=0, -105 dB. The performance drop in the case of using only the original DWA is 20.2 dB (maximum difference between the curves a) and b).

\begin{figure}
  \centering  
  \includegraphics[width=14cm]{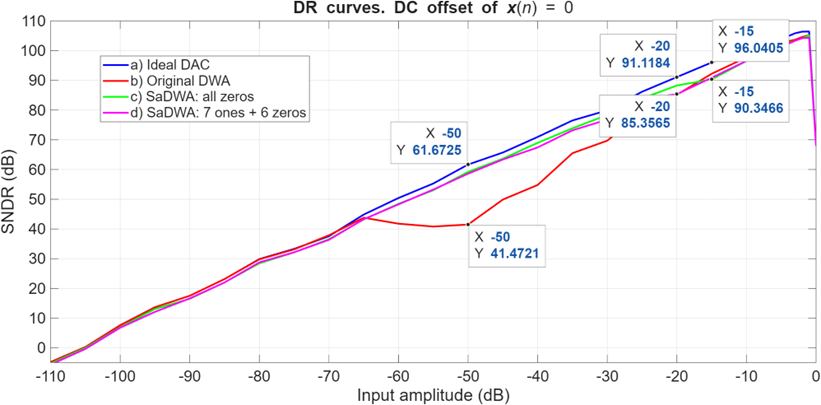}
  \caption{Dynamic range curves for different scenarios. a) Ideal DAC, b) original DWA, c) proposed SaDWA with periodic $s(n)$ and d) proposed SaDWA with $s(n) = 0$ for all $n$.}
  \label{fig_4}
\end{figure}

Although the method can be applied for different cases, the analysis will show just the  one corresponding to $s(n)$ being constant for all $n$. In particular, the case $s(n)=0$ is presented as curve d) in Fig. \ref{fig_4}. This offers better performance than the original DWA, featuring a maximum SNDR drop is 5.7 dB at the input amplitude of -20 dBfs. However, if $x(n)$ has a DC offset of $\Delta/2$ and a small AC component,  then  $y(n)$ is an almost constant sequence with predominantly a value of 4. For this case, the maximum SNDR drop is about 17 dB. 

In the particular case where $s(n) = 1$ for all $n$, the achieved SNDR is curve c) in Fig. \ref{fig_4} for a null DC offset in $x(n)$. The DC offset introduced by this $s(n)$ in the DWA block does not pass through the SDM loop in an SDM-based DAC and thus does not affect its performance. Therefore, the aforementioned 4.4 dB DR drop due to SDM quantizer overload does not occur. This is not the case in an SDM-based ADC where the DAC is in the SDM loop. Finally, if $x(n)$ had a DC offset of $-\Delta/2$ and a very small AC component, the average SDM output would be $-\Delta/2$. Then the resulting $y(n)$ is an almost constant sequence of  3 s.  Simulations in this case show a 15 dB drop in SDNR for an input amplitude of -50 dB.

\section{Conclusion}
An analysis of the mechanisms producing spurious tones in an SDM-based DAC is presented for the case where DWA is used to randomize the DAC errors. A new technique to reduce the tone level is proposed. The new method aims at overcoming the problems of previous proposals  where for certain values of DC offset in the input signal, the randomization effect is lost, and the tones reappear. The proposed solution is to add the binary sequence $s(n)$ to the DWA input. The case of constant $s(n)$ has been analyzed showing an ability to remove tones  equivalent to the previous techniques proposed while keeping the dynamic range even at high input amplitudes.


\bibliographystyle{unsrt}  
\bibliography{00mibi}

\end{document}